\documentclass{article}
\usepackage{frascatiphys,here,graphicx,subfigure}
\def\be{\begin{eqnarray} &&}
\def\nonu{\nonumber \\ &&}
\def\ee{\end{eqnarray}}
\def\psla{\slash \! \! \!}
\def\Psla{\slash \! \! \! \! }
\begin{document}
\title{
ELECTROMAGNETIC DECAYS OF VECTOR MESONS\\
IN A COVARIANT MODEL
}
\author{Silvia Pisano        \\
{\em Universit\`a di Roma "Sapienza", Rome, Italy} \\
 Tobias Frederico      \\
{\em ITA-CTA, S\~ao Jos\'e dos
Campos, S\~ao Paulo, Brazil}\\
Emanuele Pace\\
{\em Universit\`a di Roma and INFN "Tor Vergata", Rome, Italy}\\
Giovanni Salm\`e\\
{\em INFN Sezione di Roma, Rome, Italy}
}
\maketitle
\baselineskip=11.6pt
\begin{abstract}
 A fully covariant model  for describing the electromagnetic decay of  Vector
 Mesons, both in  
light and in  heavy sectors, is presented. The main ingredients of our approach
are i)
an Ansatz for the Bethe-Salpeter vertex for 
Vector Mesons, 
and ii) a Mandelstam-like formula for the electromagnetic decay constant. 
The free parameters of our approach are fixed through a comparison
with
the transverse momentum
distribution obtained within a
 Light-Front Hamiltonian Dynamics framework with constituent quarks. 
 Preliminary results for both the decays constants and the probability of
the valence component are shown.
\end{abstract}
\baselineskip=14pt
\section{Introduction}
Aim of this contribution  is to present a fully
covariant model for describing the  electromagnetic 
(em) decay of
 Vector Mesons (VM's), both in light and heavy sectors. To this end,   
 a simple analytic form for the Bethe-Salpeter 
(BS)
amplitude of VM's is adopted in order to perform without any  further 
approximation
the calculations of the decay constants. Moreover, with such an Ansatz 
one can
easily evaluate 
the so-called transverse momentum distribution of a constituent inside the
VM (see de Melo et al \cite{melo02} for the pion case), that plays an 
essential role   for fixing the value of
the parameters appearing in our approach,  and in turn for including 
some non perturbative inputs in our analytical Ansatz. A possible form of
the BS amplitude for an interacting $q\bar q$ system with $J=1$, can be 
written
as follows
\be
\Psi_\lambda(k,P)=S(k,m_1)~\left [\epsilon_\lambda(P)
\cdot V(P)\right ]~
\Lambda_{VM}(k,k-P)~S(k-P,m_2)
\ee
where $S(p,m)$ is the Dirac propagator of a constituent with mass $m$, 
 $P^\mu$ the  four-momentum of a VM
 with  mass $P^2=M^2$, $\epsilon^\mu_\lambda(P)$ its polarization
four-vector, $\lambda$ the helicity, $V^\mu(k,k-P)$ the Dirac structure of the amplitude and 
$\Lambda_{VM}(k,k-P)$ the
momentum dependence of the BS amplitude.  In particular, the adopted covariant form 
for the Dirac structure is
the  
familiar one (transverse to $P^\mu$), viz
\be V^\mu(P) ={M \over M+m_1+m_2}~\left [\gamma^\mu -{P^\mu ~\Psla P \over M^2} +\imath {1 \over M}~
\sigma^{\mu\nu} P_\nu\right ]
\ee
that in the limit of non interacting system leads to the Melosh Rotations for a
$^3S_1$ system \cite{jaus91,FPPS}, as expected. For the  present
preliminary calculations,
the momentum dependence has the following simple   
form with single poles,  viz
\be \Lambda_{VM}(k,k-P)= 
{\cal  N}~\left[k^2-m_{1}^2+(P-k)^2-m_{2}^2\right ]~\times \nonu\Pi_{i=1,3}  
{1\over \left[k^2-m_{R_i}^2 +\imath\epsilon\right ]
\left [(P-k)^2-m_{R_i}^2 +\imath\epsilon\right ]}
\label{lambda}
\ee
where $m_{R_i}$, $i=1,2,3$ are the free parameters of  our Ansatz (to be determined as
described below), ${\cal  N}$ the normalization factor, that can be derived
 by imposing the
standard normalization  for the Bethe-Salpeter amplitude, in 
Impulse Approximation (i.e. with free propagators for the
constituents). The form chosen for $\Lambda_{VM}(k,k-P)$ allows one both to implement   
 the correct symmetry under the exchange of the quark momenta
(for equal mass constituents) and to avoid
 any free propagation of the constituents (cf the numerator in 
 Eq. (\ref{lambda})).

To determine $m_{R_i}$ in Eq. (\ref{lambda}),  we first define the 
constituent transverse momentum
distribution inside the VM, $n( k_{\perp})$, along the same guidelines adopted 
by de Melo 
et al. \cite{melo02} 
for the
pion, within a Light-Front Hamiltonian Dynamics approach. 
In a frame where ${\bf P}_\perp={\bf 0}$, one has
 \be
 n(k_{\perp})={N_c\over P_{q\bar q}(2\pi)^3\left [P^+\right]^2}
\int_0^{2\pi} d\theta_{\hat k_\perp}\int_0^1 {d\xi ~M^2_0\over
\xi (1-\xi)} ~ 
|\Phi(\xi,{\bf k_{\perp}};m_{R_i})|^2
 \label{nk}\ee
where $N_c$ is the number of colors,  $k_\perp=|{\bf k_{\perp}}|$ and 
$\Phi(\xi,{\bf k_{\perp}};
m_{R_i})$ is the
valence wave function associated to a given BS amplitude, see, e.g., Huang and 
Karmanov\cite{HK} and Frederico et al\cite{FPPS}.
  In Eq. (\ref{nk}), the probability $P_{q\bar q}$ of the valence
component  reads
\be
P_{q\bar q}=
{N_c\over (2\pi)^3\left [P^+\right]^2}
\int_0^1 {d\xi \over
\xi ~(1-\xi)}\int d{\bf k}_\perp M^2_0~|\Phi(\xi,{\bf k_{\perp}};
m_{R_i})|^2
\ee
Finally, $n( k_{\perp})$ is normalized as: 
$
\int k_\perp~d{ k}_\perp ~n( k_{\perp})=1
$.

 In spite
of the simple form assumed for $\Lambda_{VM}$, one can nicely fit the
constituent transverse 
momentum distributions 
 obtained within 3-D approaches, that i) retain only the valence
component of the VM's and ii) are able to yield a reasonable description of the
spectrum. In this work we have extracted the parameters $m_{Ri}$ in 
Eq. (\ref{nk}) by fitting 
$n(k_{\perp})$   to the
corresponding quantity obtained from i) a Harmonic Oscillator model 
(see, e.g. Figs. 1 and 2), ii)  the
Godfrey-Isgur model \cite{GI} and iii)  an adapted version of the model by  
Salcedo 
et al \cite{qcd_tob} (ITA model).

\section{The Mandelstam formula for em decay constant}
In order to evaluate the em decays constants, $f_V$, we adopted 
a Mandelstam-like formula\cite{mandel} (see also de Melo et al.\cite{pionprd})). 
The starting point is the {\em macroscopic} definition of $f_V$, through 
the transition matrix element of the em current for a given neutral  VM, viz
\be
\langle 0|J^{\mu}(0)| P,\lambda \rangle =
\imath \sqrt{2}f_V \epsilon_{\lambda}^{\mu}
\label{eq:fv} 
\ee
The decay constant $f_V$ is related to the em decay width 
as follows
\be
\Gamma_{e^+ e^-}=\frac{8\pi\alpha^2}{3}\frac{|f_V|^2}{M^3}
\label{eq:gamma} 
\ee
In our model, the transition matrix element in Eq. (\ref{eq:fv}) can be 
 approximated {\it microscopically}
{\it \`a la} Mandelstam through
\be
\langle 0|J^{\mu}(0)| P,\lambda \rangle=
{\cal F}_{VM}~{{N_c}{\cal N} \over (2\pi)^4}\int d^4 k
{\Lambda_{VM}(k,k-P,m_1,m_2)\over 
  (k^2-m^2_1+ \imath \epsilon)~[(P-k)^2-m^2_2
  + \imath \epsilon]} \times \nonu
{\rm Tr} [ \epsilon_{\lambda}(P) \cdot V(P) ~(\psla k -\psla P +m_2)\gamma^\mu
(\psla k+m_1)]
\label{eq:mandel} 
\ee 
where $$ {\cal F}_\rho= {(Q_u-Q_d) \over \sqrt{2}} \quad \quad
{\cal F}_\phi= Q_s\quad \quad
{\cal F}_{J/\Psi}= Q_c
$$
with $Q_i$ the quark charge.
In Tabs. \ref{tab:hosci}, \ref{tab:gi}, \ref{tab:qcd-in}, the preliminary
results for  both
valence probability, $P_{q\bar q}$, and   em decay widths, 
$\Gamma_{e^+ e^-}$  are shown. 
Even if a more refined evaluations are in progress, some comments 
 are in order: i) for the Harmonic Oscillator
model the light meson decay widths can be reasonably well described 
 (in the light sector the confining interaction is quite relevant),
while the $J/\psi$  one is largely underestimated; ii) for the 
Godfrey-Isgur model \cite{GI}, the heavy sector is well reproduced, 
while the light sector is overestimated, and this appears correlated 
to the poor
estimate of the valence probability (work in progress suggests that an 
 Ansatz for the BS amplitude with a more rich structure  substantially improves
 the comparison); iii) 
for the adapted version of 
the ITA model \cite{qcd_tob} the same pattern of the Harmonic Oscillator
case has been found, even if more dynamical contents are present in this 
model.
%
%
%
%
\begin{table}
\centering
\caption{\textsl{Preliminary VM em decay widths within the 
    {{ Harmonic Oscillator}} model. Adopted quark masses: $ m_u=$ 0.310 GeV,
$m_s=$ 0.460 GeV,
$m_c=$ 1.749 GeV,
$m_b=$ 5.068 GeV,}
\label{tab:hosci}}

\begin{tabular}{|c|c|c|c|c|c|}\hline
{VM} & {$m_{VM}$ (MeV)} &{$P_{q\bar{q}}$} & {$\Gamma_{e^+ e^-}^{th}$  (keV)}& {$\Gamma_{e^+ e^-}^{exp}$  (keV)}\\
\hline
\hline
$\rho$ &775.5 $\pm$ 0.4  &  0.884 & 10.328  & 7.02 $\pm$ 0.11  
\\  \hline
$\phi$ & 1019.460 $\pm$ 0.019 & 0.961 &1.582  & 1.32 $\pm$ 0.06 
\\  \hline
$J/\psi$ & 3096.916 $\pm$ 0.011 &0.787& 1.572 &5.55 $\pm$ 0.14   
\\  \hline
\end{tabular}
 
\end{table}
%
%
%
%
%
\begin{table}
\centering
\caption{\textsl{Preliminary VM em decay widths within the  
Godfrey-Isgur\cite{GI} model, $m_u=$ 0.220 GeV,
$m_s=$ 0.419 GeV,
$m_c=$ 1.628 GeV,
$m_b=$ 4.977 GeV.}
\label{tab:gi}}

\begin{tabular}{|c|c|c|c|c|c|}\hline
{VM} &{$m_{VM}$ (MeV)}&{$P_{q\bar{q}}$} &{$\Gamma_{e^+ e^-}^{th}$ (KeV)} &{$\Gamma_{e^+ e^-}^{exp}$ (keV)} \\
\hline
\hline
$\rho$ &775.5 $\pm$ 0.4  & 0.411& 18.098 & 7.02 $\pm$ 0.11 
\\  \hline
$\phi$ & 1019.460 $\pm$ 0.019 & 0.906 &3.733 &1.32 $\pm$ 0.06 
\\  \hline
$J/\psi$ & 3096.916 $\pm$ 0.011&0.908 & 5.911&5.55 $\pm$ 0.14 
\\  \hline
\end{tabular}
 
\end{table}

\begin{table}
\centering
\caption{\textsl{Preliminary VM em decay widths within an adapted version of the ITA model.
   \cite{qcd_tob}. Adopted quark masses:
$m_u=$ 0.334 GeV,
$m_s=$ 0.460 GeV,
$m_c=$ 1.791 GeV,
$m_b=$ 4.679 GeV}.
\label{tab:qcd-in}} 

\begin{tabular}{|c|c|c|c|c|c|}\hline
{VM} &{$m_{VM}$ (MeV)} &{$P_{q\bar{q}}$} &{$\Gamma_{e^+ e^-}^{th}$  (keV)}&{$\Gamma_{e^+ e^-}^{exp}$ (keV)}\\
\hline
\hline
$\rho$ &775.5 $\pm$ 0.4   & 0.913 & 7.548  & 7.02 $\pm$ 0.11 
\\  \hline
$\phi$ & 1019.460 $\pm$ 0.019  & 0.995&1.294  &1.32 $\pm$ 0.06 
\\  \hline
$J/\psi$ & 3096.916 $\pm$ 0.011 &0.726& 1.250 &5.55 $\pm$ 0.14  
\\  \hline
\end{tabular} 
\end{table}
\begin{figure}
    \begin{center}
        {\includegraphics[width=5.5cm]{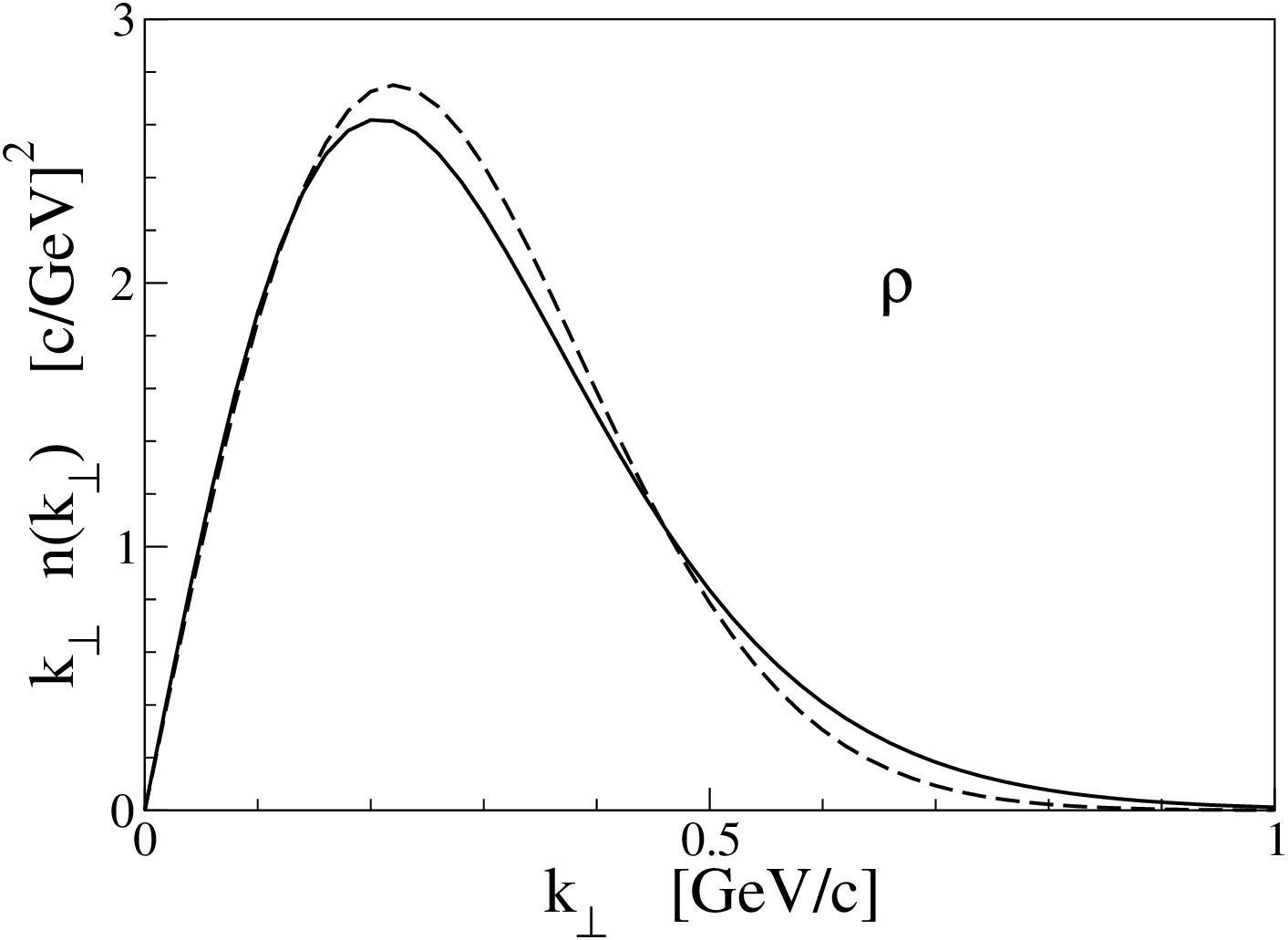}}
\hspace{.5cm} 
        {\includegraphics[width=5.5cm]{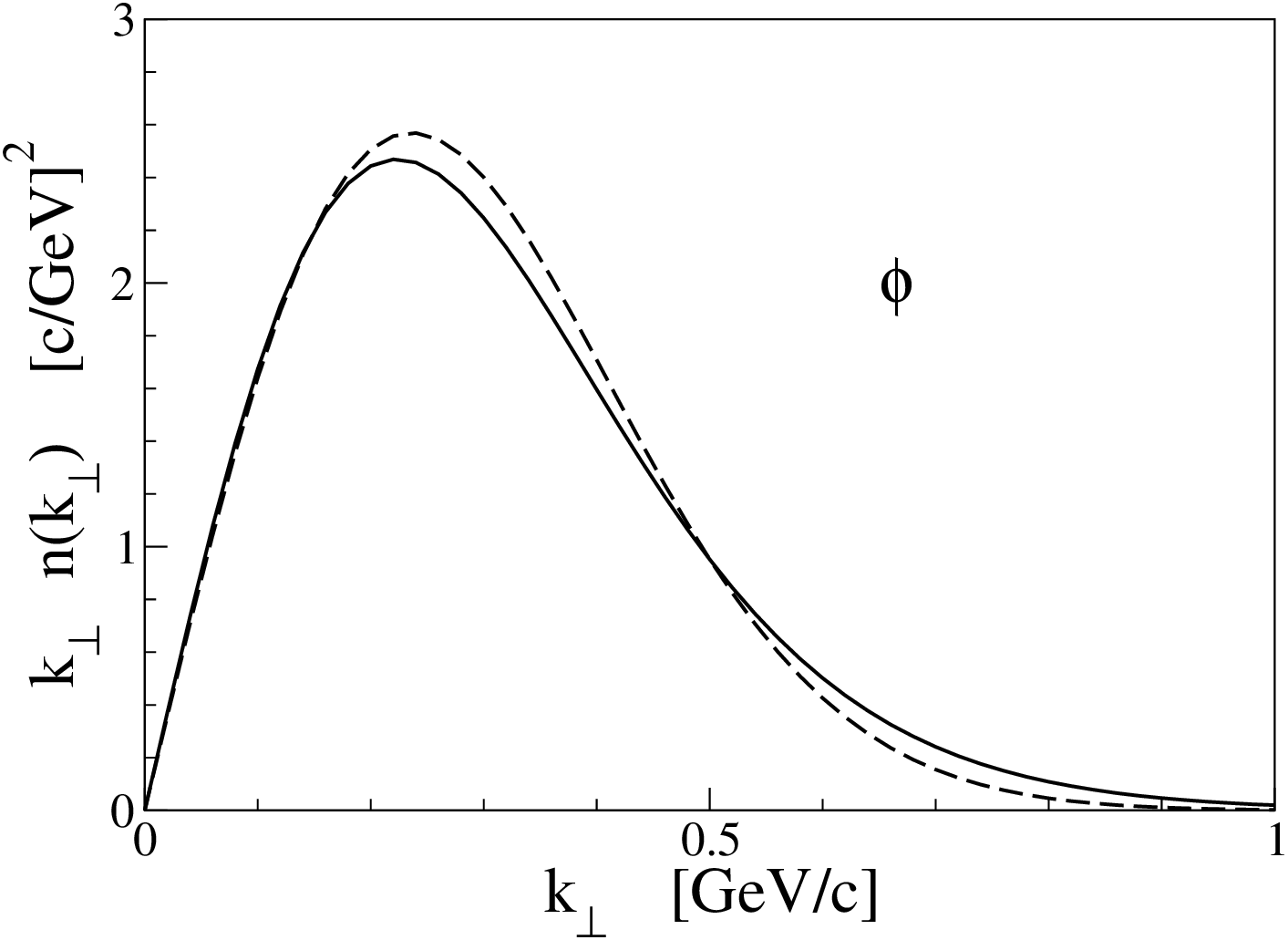}}
\caption{\it Transverse momentum distributions for a constituent inside  
$\rho$ and $\phi$ 
  vs the quark transverse momentum. Dashed line:  Harmonic 
 Oscillator model. Solid line: fit  by using the analytic Ansatz in Eq. (\ref{lambda})}
    \end{center}
\end{figure}
\begin{figure}
    \begin{center}
        {\includegraphics[width=5.5cm]{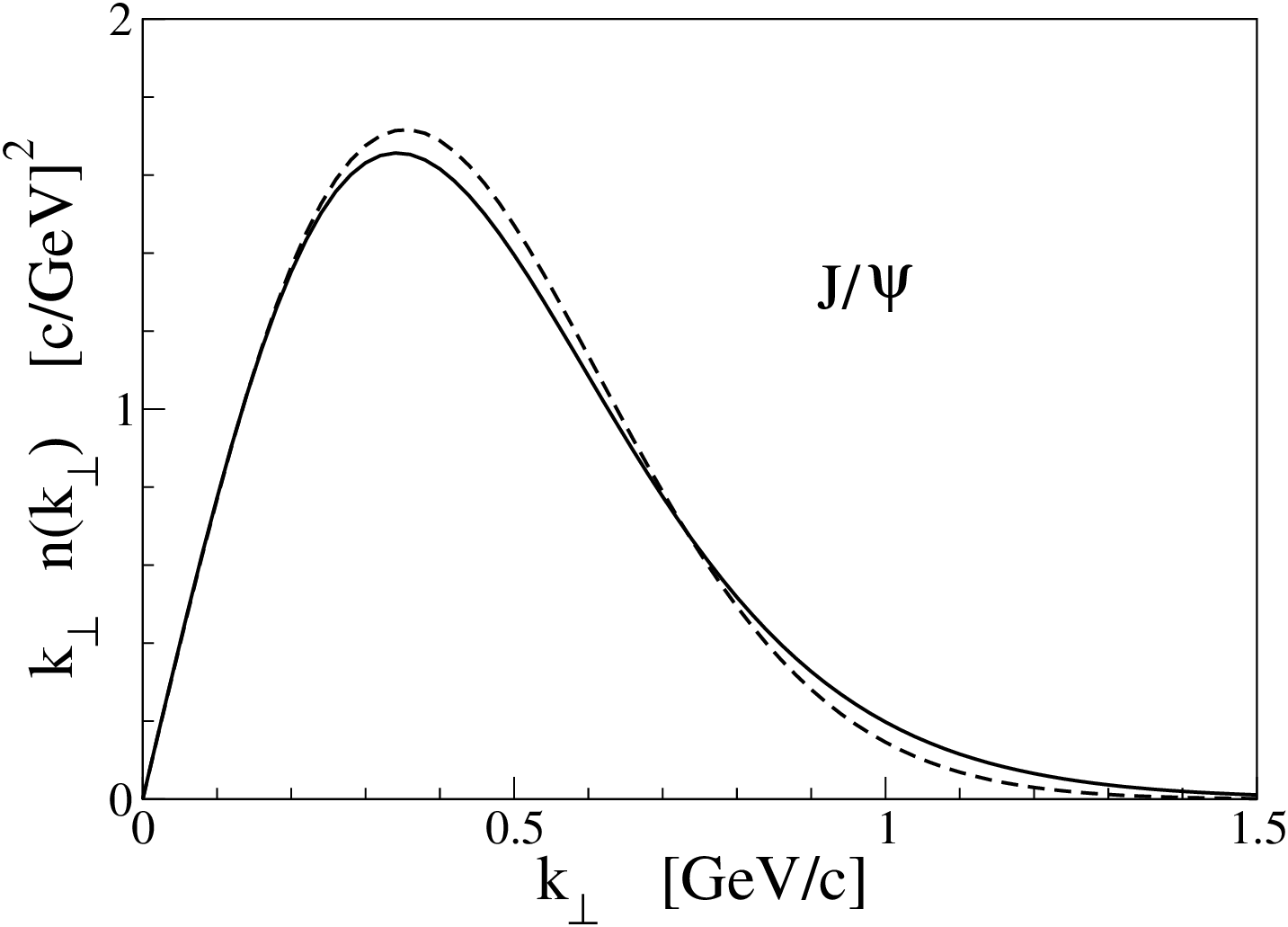}}
        \caption{\it The same as in Fig. 1, but for  $J/\psi$.}
    \end{center}
\end{figure}

\section{Conclusions}
In this contribution we have presented the main ingredients of our model for
evaluating
both the em
decay width of the ground states of VM's and the probability of the valence
component of the state. In our fully covariant model, a simple, analytic Ansatz
for the BS amplitude is proposed, and the three parameters, $m_{Ri}$, for each 
neutral VM,
are determined through a fitting procedure, based on the
transverse momentum distribution of a constituent inside a given VM, obtained
within a Light-Front Hamiltonian Dynamics framework.  
From this first  comparisons between our results  and  the experimental data, 
one could argue
that two different regimes occur in the light ($\rho$, $\phi$) and in the 
heavy sector ($J/\psi$).
From the theoretical side, indeed, the Harmonic Oscillator and the adapted 
ITA\cite{qcd_tob} models
 seem 
to better reproduce
the light mesons (cf Tab. \ref{tab:hosci}, \ref{tab:gi}) while for 
the heavy sector the
Godfrey-Isgur model\cite{GI} seems to work better (cf. Tab. \ref{tab:qcd-in}).

The work in progress will substantially improve the present calculations,
in two respect: both introducing a more refined Ansatz for the BS amplitude and extending our
investigation to the em decay of the $\Upsilon$.
\end{document}